\documentclass{aa}  

\usepackage{graphicx}
\usepackage{txfonts}
\newcommand{\mytilde}{\raise.17ex\hbox{$\scriptstyle\mathtt{\sim}$}}

\begin{document}

   \title{K2 Variable Catalogue: Variable Stars and Eclipsing Binaries in K2 Campaigns 1 and 0}

   \subtitle{}

   \author{D. J. Armstrong
          \and
          J. Kirk \and K. W. F. Lam \and J. McCormac \and S. R. Walker \and D. J. A. Brown \and H. P. Osborn \and D. L. Pollacco \and J. Spake
          }

   \institute{Department of Physics, University of Warwick,
              Gibbet Hill Road, Coventry, CV4 7AL, UK\\
              \email{d.j.armstrong@warwick.ac.uk}
             }

   \date{Received ; accepted}

 
  \abstract
   {}
   {We have created a catalogue of variable stars found from a search of the publicly available K2 mission data from Campaigns 1 and 0. This catalogue provides the identifiers of 8395 variable stars, including 199 candidate eclipsing binaries with periods up to 60d and 3871 periodic or quasi-periodic objects, with periods up to 20d for Campaign 1 and 15d for Campaign 0.}
   {Lightcurves are extracted and detrended from the available data. These are searched using a combination of algorithmic and human classification, leading to a classifier for each object as an eclipsing binary, sinusoidal periodic, quasi periodic, or aperiodic variable. The source of the variability is not identified, but could arise in the non-eclipsing binary cases from pulsation or stellar activity. Each object is cross-matched against variable star related guest observer proposals to the K2 mission, which specifies the variable type in some cases. The detrended lightcurves are also compared to lightcurves currently publicly available.}
   {The resulting catalogue is made available online at http://deneb.astro.warwick.ac.uk/phrlbj/k2varcat/, and gives the ID, type, period, semi-amplitude and range of the variation seen. We also make available the detrended lightcurves for each object.}
   {}


   \maketitle
%

\section{Introduction}
The K2 mission \citep{Howell:2014ju} is the survey now being conducted with the repurposed Kepler space telescope, and became fully operational in June 2014. It will survey a series of fields near the ecliptic, returning continuous high-precision data over an 80 day period for each field. Despite the reaction wheel losses that ended the Kepler prime mission, K2 has been estimated to be capable of 80ppm precision for V=12 stars, close to the sensitivity of the primary mission. All data will be public, although at the time of writing only campaigns 0 and 1 have been released, in September and December 2014. As the mission progresses, much more data should become available. Targets are provided by the Ecliptic Plane Input Catalogue (EPIC) which is hosted at the Mikulski Archive for Space Telescopes (MAST) along with the available data products. Approximately 7500 objects were observed during Campaign 0 and \mytilde22000 during Campaign 1, mostly in `long-cadence' (a cadence of \mytilde 30 min). A few (13 and 56 respectively) were also observed in `short-cadence' (\mytilde 1 min). All identification processes in this catalogue were performed on the long cadence dataset. A number of objects located near (the specific distance varies, but is of order a few tens of arcseconds) these EPIC targets were also observed but are not in the EPIC catalogue. These were not used in making this catalogue.

The K2 mission will be of great use to a wide range of astronomical research areas. Although the original Kepler space telescope was primarily aimed at the detection and study of exoplanets, its high precision lightcurves were used for studies with astroseismology \citep[e.g.][]{Chaplin:2013jf}, stellar rotation \citep[e.g.][]{Reinhold:2013iz} and eclipsing binaries \citep[e.g.][]{Prsa:2011dx}, to name just a few. Already the K2 mission has been used to identify new candidate eclipsing binaries \citep{Conroy:2014gy}, and produced new interesting planetary systems \citep{Crossfield:2015vm,Vanderburg:2014wi}. The utility of Kepler extended to the study of variable stars, with a number of studies en masse and individually of different kinds of variable \citep[e.g.][]{McQuillan:2012dj,Holdsworth:2014hc,Stello:2014is,Banyai:2013hu}. Catalogues were made available using a variety of techniques \citep{Debosscher:2011kz,Uytterhoeven:2011jv}. Recently such catalogues have begun appearing for the K2 mission, including a recent cross match with the TESS target catalogue \citep{Stassun:2014wz}. There are also studies ongoing of variable stars within the K2 fields of view, such as that of \citet{Nardiello:2015ev}, where variable stars within two open clusters were identified by ground based photometry.

After the Campaign 0 data became available a preliminary version of this catalogue was made available \citep{Armstrong:2014wf}, identifying and classifying stars showing variability in the K2 observations. Here we formally release that catalogue, as well as including the Campaign 1 data and adding eclipsing binaries from both campaigns. 


\section{Data Preparation}
\label{sectdataprep}
\subsection{Data Source and Extraction}
\label{sectextract}
Our lightcurves were obtained from the MAST archive of K2 data (Campaign 1: Data Release 1, Campaign 0: Data Release 2). These are at present only available as Target Pixel Files, giving the pixel time series of a variably sized window surrounding the proposed target. At this stage we used only the long cadence observations (bearing in mind that each short cadence target also has data in long cadence). We also limit ourselves to objects classified by MAST as `STARS' or `EXTENDED SOURCES', ignoring observations otherwise classified (these include clusters, comets, and other targets). Work on variability within K2 clusters has recently been carried out by \citet{Nardiello:2015ev}. For each entry in the EPIC catalogue which we considered, one lightcurve was produced. This means that other objects near the planned targets, which were observed by K2 but not explicitly in the EPIC catalogue, were not considered here. The data were cut to exclude regions at the start of each campaign due to course point and safe mode events. For Campaign 0, data before 1940.5 (BJD-2454833, as found in the MAST data) were removed, leaving a baseline of \mytilde32 days. For Campaign 1, data before 1978.5 (BJD-2454833) were removed leaving \mytilde79 days. The removed points were not reincluded at a later stage. 

\begin{figure}
\resizebox{\hsize}{!}{\includegraphics{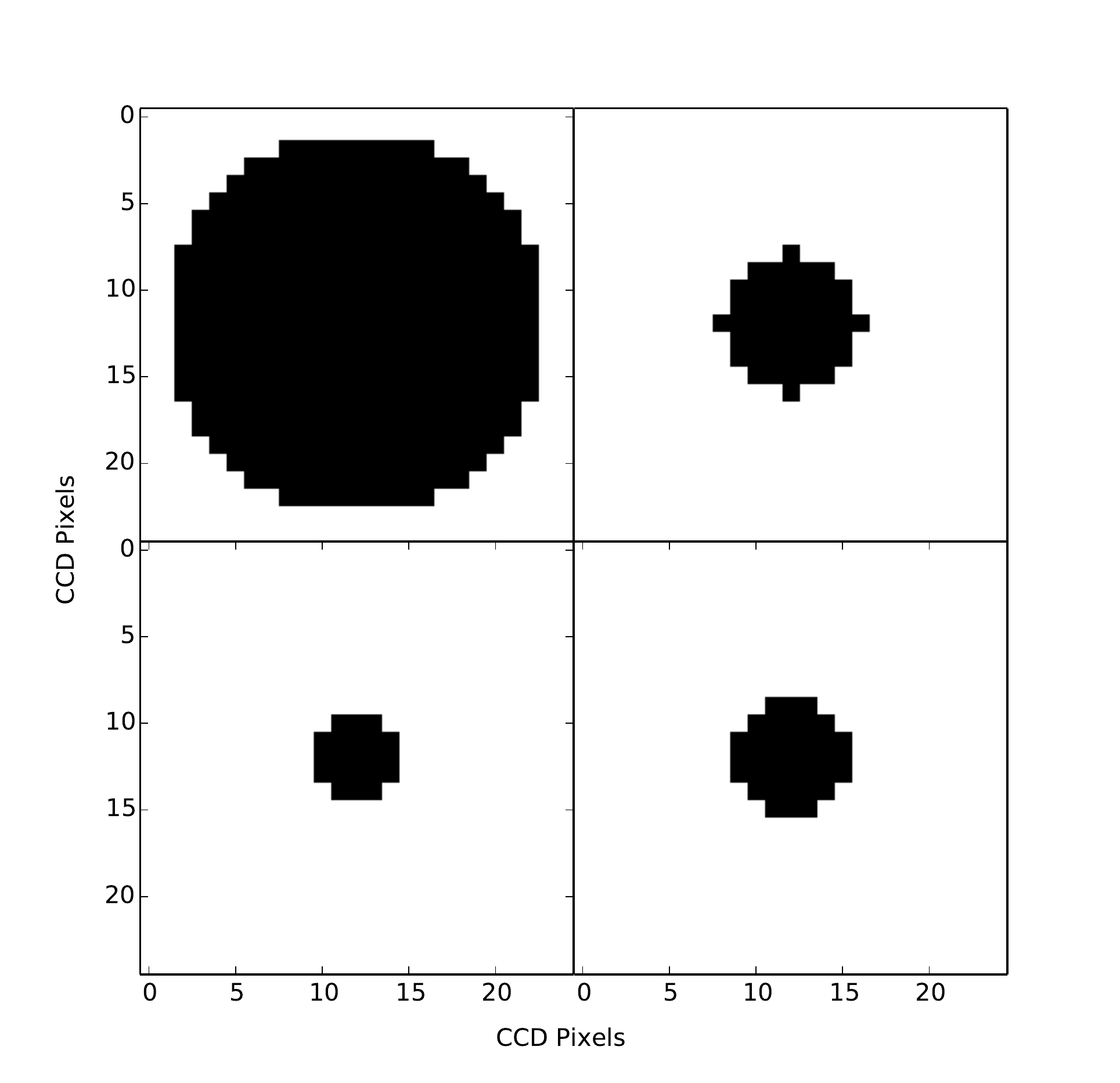}}
\caption{Example apertures for (clockwise from top left) Kepler magnitude $<=10$, $<=13$, $<=16$, $>16$.}
\label{figapertures}
\end{figure}

We developed a program to allow more flexible extraction according to the needs of K2 \citep[as in for example][although our extraction is more simple]{Aigrain:2015ew}. The WCS information contained within the target pixel files was utilised to find the central pixel of the target (we found the WCS information to generally be accurate to within 1 pixel.). An aperture was then set depending on the brightness of the target. We found through trial and error that apertures of radius 3, 4, 5 and 12 pixels, for targets with Kepler magnitude $>16$, $<=16$, $<=13$, $<=10$ respectively, produced good results while minimising the chance of blending with other targets in the window (see Figure \ref{figapertures} for example apertures, within which each pixel is given full weighting). The aperture was recentred to the brightest pixel within its initial position derived from the WCS coordinates (using the median brightness of each pixel measured over the whole dataset). Apertures for objects with Kepler magnitudes $<=10$ were made particularly large due to the bleeding effect which can occur for these targets, and which covers large numbers of pixels. We limited ourselves to 4 aperture sizes to allow easy recreation of the aperture when checking data without looking into the detail of the files. The relation of target magnitude to apparent size on the CCD is also not trivial, and can vary even for objects of the same magnitude. Hence a smaller number of fixed (larger) apertures avoid systematic issues that may be introduced by assuming a tight magnitude-aperture size relation and for example letting the aperture size vary smoothly with magnitude. It is possible to recreate the used apertures by using the new header card 'AP\_RAD' provided in the data files (see Table \ref{tabnewheaders}). This is the squared aperture radius, and a pixel is within the aperture if $(X_{pixel}-X_{centre})^2 + (Y_{pixel}-Y_{centre})^2 <$ AP\_RAD, where X and Y are pixel coordinates in each axis. Once a raw lightcurve was available, background subtraction was performed using a background value determined by the median value of pixels outside the aperture at each timestamp. Although a simple method, we found that this was generally robust. The use of the median avoids significant bias by other sources except in a small number of cases, especially as we do not consider cluster observations. The error on the background determination was found from the median of the absolute deviation from their median of the out of aperture pixels, known as the `Median Absolute Deviation'. This was then added in quadrature along with the pixel errors inside the aperture to produce the extracted flux errors.

At \mytilde2016 (BJD-2454833) during the Campaign 1 data, the spacecraft pointing changed significantly, resulting in movement of targets by over a pixel in some cases. As such we recalculated the aperture centres after this time, using the same aperture shape.

\subsection{Data Detrending}
\label{sectdetrend}
The main source of systematic noise in K2 data is pointing drift, as has been pointed out previously \citep{Vanderburg:2014bi}. This has been claimed to be from either pixel-to-pixel flat fielding errors or a combination of aperture losses and source crowding. We independently designed and implemented a method similar to that proposed by \citet{Vanderburg:2014bi} in order to detrend our lightcurves, which removes all noise correlated with the pointing drift regardless of its source.

The row and column centroid positions were calculated for each timestamp. This was done through the relation

\begin{equation}
\phi_{x} = \frac{\sum\limits_{x=0}^{n_x} \left(x\sum\limits_{y=0}^{n_y}z(x,y)\right) }{\sum\limits_{x=0}^{n_x} \sum\limits_{y=0}^{n_y} z(x,y) }
\end{equation}

where z(x,y) is the flux at the pixel in row x and column y, $n_x$ is the total number of pixels in each row, $n_y$ the total number of pixels in each column, and $\phi_x$ the resulting row centroid. The column centroid is calculated by changing each x to a y and vice versa.

\begin{figure}
\resizebox{\hsize}{!}{\includegraphics{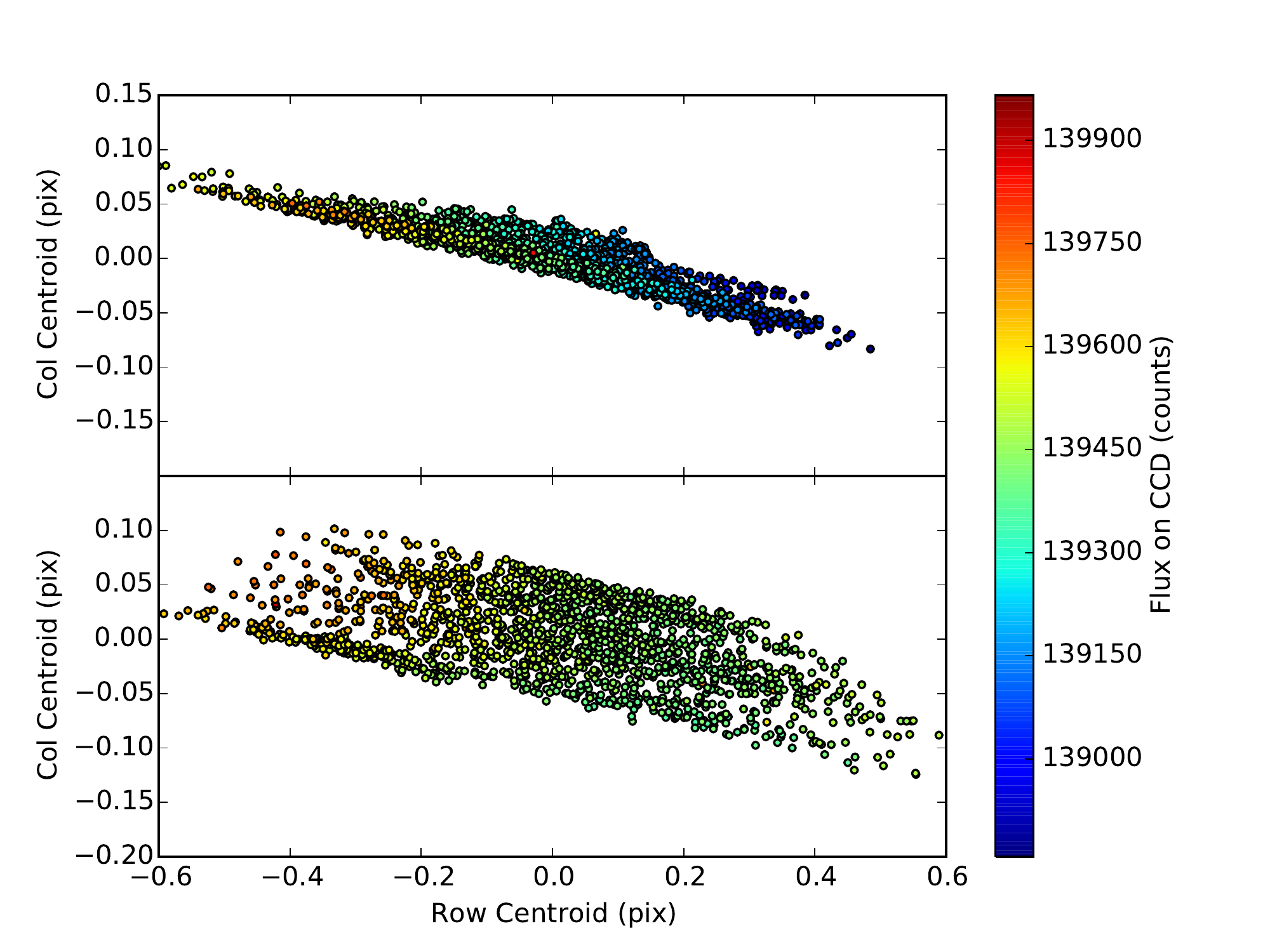}}
\caption{Surface of raw flux against CCD position for EPIC201552917. The flux is split at 2016 (BJD-2454833), with the first segment above and the second segment below. Centroids have been offset to their respective means.}
\label{figscatter}
\end{figure}

\begin{figure}
\resizebox{\hsize}{!}{\includegraphics{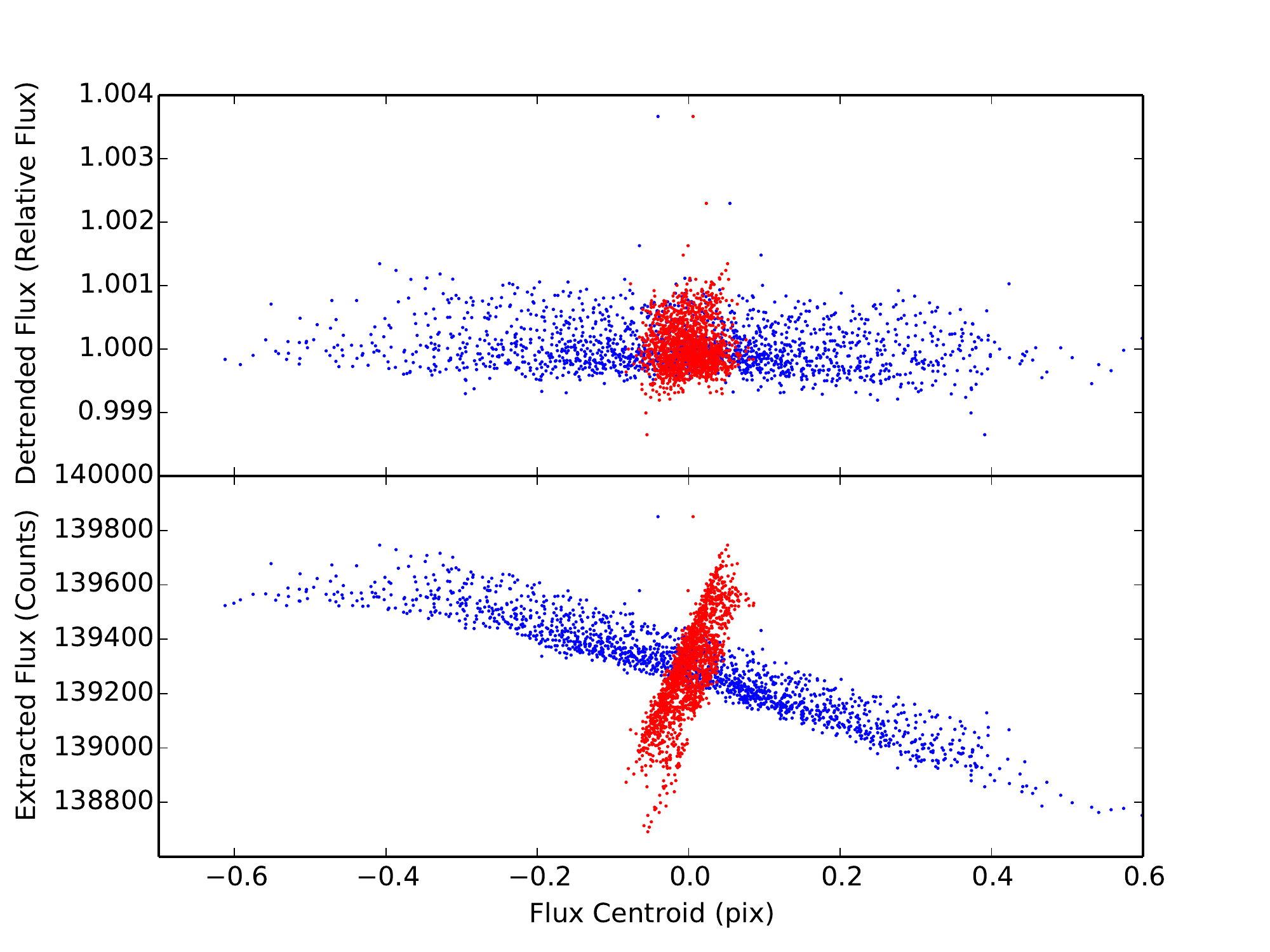}}
\caption{Correlation of flux to CCD centroid position for the extracted (bottom) and detrended (top) lightcurves of EPIC201552917. Row centroid is shown in blue, column centroid in red.}
\label{figdecorrelate}
\end{figure}

\begin{figure}
\resizebox{\hsize}{!}{\includegraphics{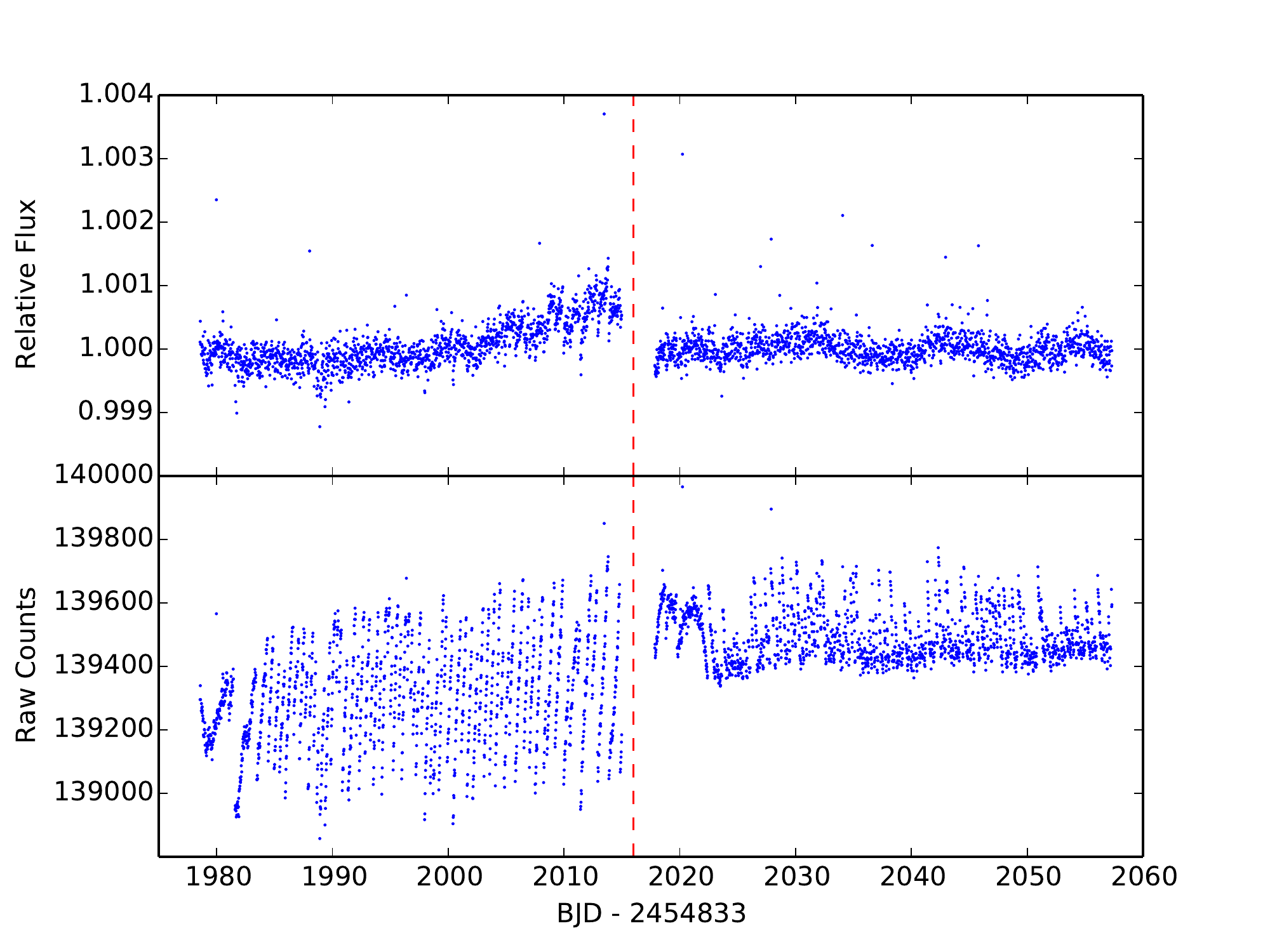}}
\caption{Extracted (bottom) and Detrended (top) lightcurves for EPIC201552917, showing some systematic noise. The dashed line indicates the time where detrending was split.}
\label{figdetflat}
\end{figure}

\begin{figure}
\resizebox{\hsize}{!}{\includegraphics{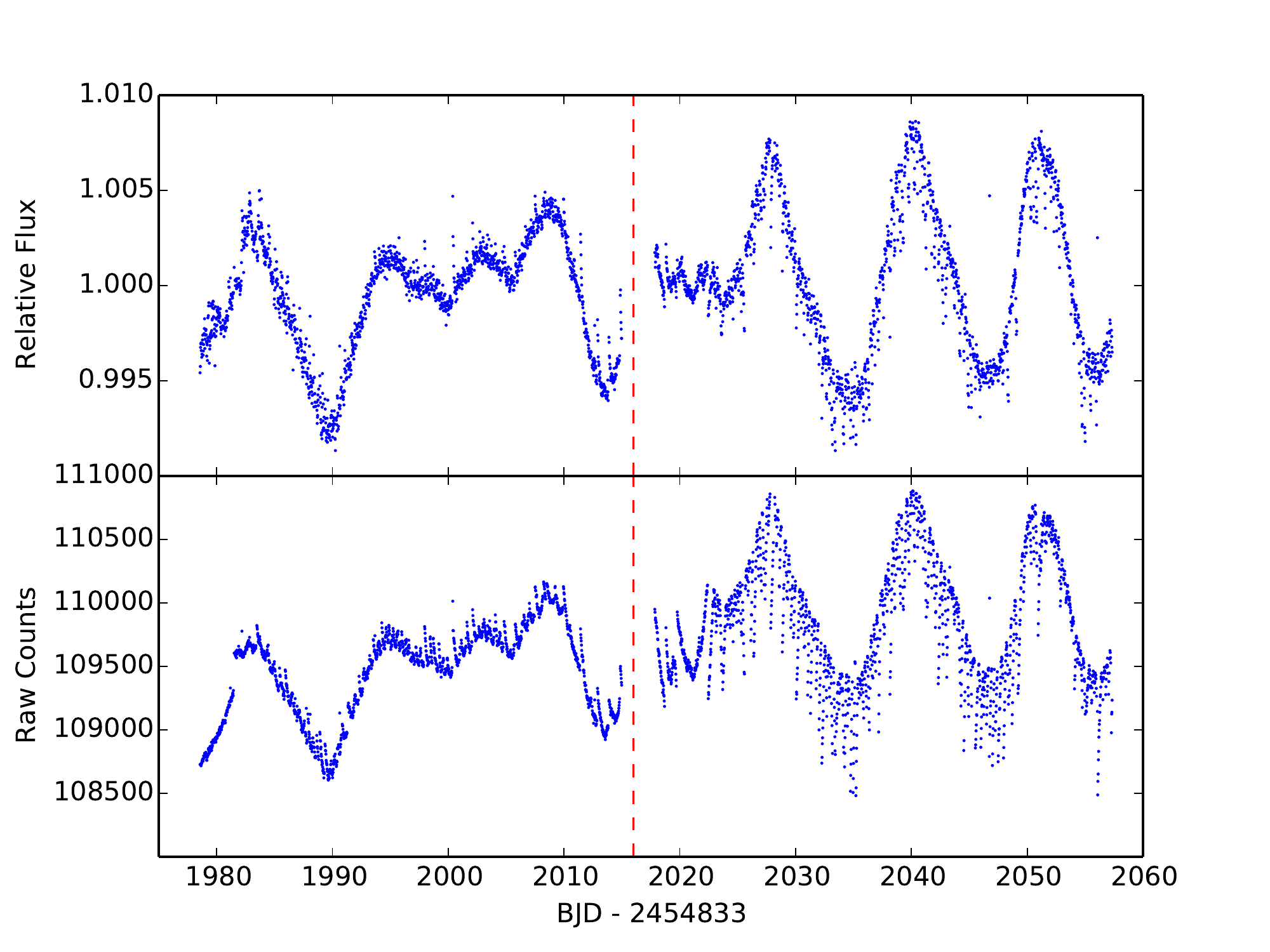}}
\caption{Extracted (bottom) and Detrended (top) lightcurves for EPIC201809540, showing quasi periodic variability. The dashed line indicates the time where detrending was split.}
\label{figdetvar}
\end{figure}

At this stage points near a thruster firing event were cut, detected as those to either side of times where the point-to-point centroid shift was greater than 3 times the median point-to-point shift across the dataset. The centroid positions were then used to create a 2D surface of raw flux against position of the centroid on the CCD. An example such surface is shown in Figure \ref{figscatter}. If the pointing drift had no impact on the flux, this surface should show no correlation. Instead, in the majority of cases a strong trend was seen. This trend was identified through binning the data into 10 evenly spaced bins in row and 10 in column, making 100 individual bins in total. The median flux in each bin was then taken and interpolated between linearly using the SciPy griddata function\footnote{http://www.scipy.org/, v0.15} \citep{scipyref}, creating a smooth surface mapping the variation caused by the observed centroid shifts. We used SciPy as it provides a versatile analysis tool for scientific work in Python. Bins containing fewer than 3 points were cut and not used for interpolation. The resulting surface was divided out, decorrelating the flux from spacecraft pointing and providing a lightcurve in flux relative to unity. The griddata function can ignore some points if they have values inconsistent with the surface formed by the majority of input points; see \citet{Barber96thequickhull} for a full description of the QHull algorithm, which forms the basis of griddata when used linearly and is more complex than can be concisely explained here. The surface at such points was defaulted to the nearest valid bin value. The correlation of an example lightcurve with centroid position is shown in Figure \ref{figdecorrelate}, both before and after detrending. In addition, outliers were removed by cutting data points where the centroid position was greater than 5 times the median distance from the median centroid position across the dataset. In all these situations medians rather than means and standard deviations were used in order to avoid the effects of large outliers. Example lightcurves, pre and post detrend, are shown in Figures \ref{figdetflat} and \ref{figdetvar}. We note that in Campaign 1 in particular, some systematic noise remains after detrending, likely arising from instrumental effects as seen in the original Kepler data. Some such variation can be seen in the detrended light curve of Figure \ref{figdetflat}. Such variations can be seen in the Eigen lightcurves of \citet{ForemanMackey:2015vi}. The specific origin of the variations in K2 is at present poorly understood. We have not attempted to remove this variation in this work, as it does not correlate with the pointing drift.

At the same time as the previously mentioned pointing shift at \mytilde2016 (BJD-2454833), the characteristics of the thruster firing and associated spacecraft motion also changed. We do not know the underlying reason for this and so do not provide further detail. However, we adjusted for this effect by detrending the Campaign 1 lightcurves before and after the split separately. This provided significantly improved lightcurves over results tried without a split, but has the disadvantage that long period variability can be removed. There was no need to perform such a split in Campaign 0, which contained no such characteristic change. We found that the above method worked well in most cases, but it has the weakness that intrinsic stellar variability which occurs on a similar timescale to the dataset can be removed, if the spacecraft drift spuriously correlates with it. Detrending the Campaign 1 data in two segments means that this applies to variability on a shorter timescale, of order 35 days rather than the full dataset length of 79 days. We also note that large amplitude variability which dominates over the pointing noise can also be reduced in amplitude, should it correlate with the drift. The catalogue web pages show both extracted and detrended flux, which will make such a reduction or blurring of a real signal evident if it has occurred.

There is also a beneficial side effect of this method - it automatically weakens signals associated with variability on a background blended object, as such variability can cause centroid position changes correlated with the change in flux. This applies equally to stellar variability or to background blended eclipses.

\subsection{Performance}
\begin{figure}
\resizebox{\hsize}{!}{\includegraphics{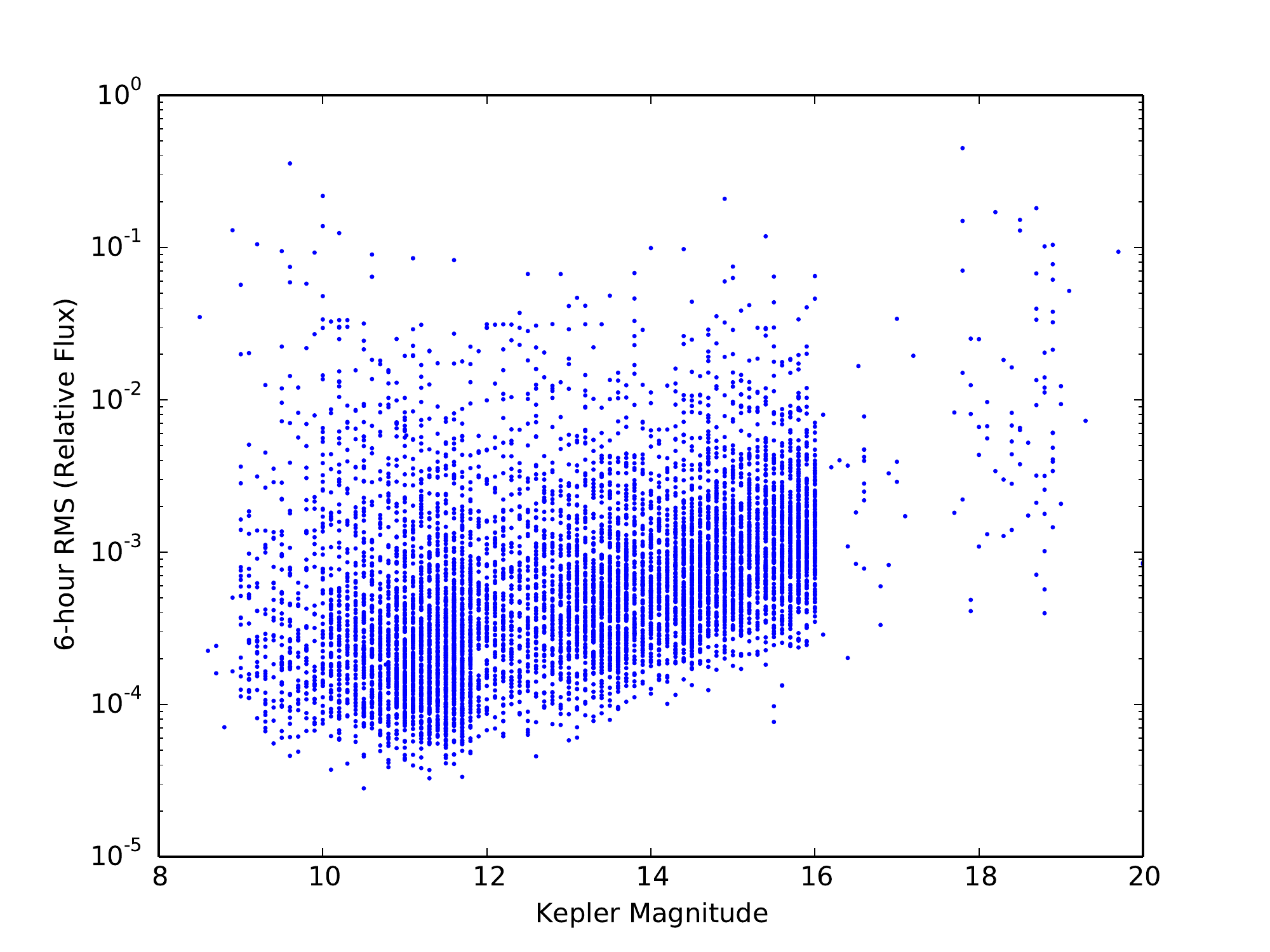}}
\caption{Root Median Square plot of all (i.e. variable and non-variable) Campaign 0 detrended lightcurves binned into 6-hour bins. A small random noise on the magnitudes has been added for clarity.}
\label{figRMS6hour}
\end{figure}

Having applied this detrending method to our set of K2 lightcurves we were in a position to test it's overall performance as compared to the other methods available for K2 data \citep[e.g.][]{Aigrain:2015ew,Vanderburg:2014bi}. For this purpose we have created a Root Median Square (RMS) plot, shown in Figure \ref{figRMS6hour}. This shows the 6-hour performance of all detrended lightcurves from Campaign 0. We limited this test to Campaign 0 as the other available detrending methods for K2 data had only released up to at most Campaign 0 at the time of writing. Magnitudes are Kepler magnitudes and were taken from the 'KEPMAG' header found within each data file. RMS values were calculated as $\textrm{RMS} = \left[\textrm{median}\left((x - \textrm{median}(x))^2\right)\right]^{\frac{1}{2}}$, where x represents the array of 6-hour binned flux values.

The plot shows a number of interesting characteristics. In particular is the slight turn up at the bright end, which is a result of the bleeding that can occur for brighter targets. In these cases it is likely that some flux was lost from the aperture. The distribution of magnitudes seen is largely a result of which proposed targets were selected for download from the spacecraft. In overall terms, the median 6-hour RMS value for our Campaign 0 detrended lightcurves was $5.39x10^{-4}$, with a `best' RMS of $2.81x10^{-5}$. We downloaded the public Campaign 0 data from \citet{Vanderburg:2014bi} to compare this result. The comparison was limited to lightcurves found in both sets of lightcurves (7691 in total). We cut points marked as being within thruster firing events, but otherwise leave the lightcurves as they are presented. Although this means that the comparison is not on precisely the same data points, it is instead between the lightcurves generated and published in both cases. As such it is a comparison of the lightcurves available, not the specific method. The median 6-hour RMS value for the \citet{Vanderburg:2014bi} lightcurves was $7.46x10^{-4}$, with a `best' RMS of $3.52x10^{-5}$, implying that our method is performing comparatively. We are aware of one other published method for K2 data detrending, that of \citet{Aigrain:2015ew}. While Campaign 0 data were not available for this method at the time of writing, the above RMS values are comparable with the results shown by that method for the K2 Engineering dataset. 
 
There are significant methodological differences between these and all other K2 detrending methods, in for example the aperture sizes and shapes used and methods of lightcurve extraction as well as the detrending itself. As such, we explicitly do not claim that our method is `better' or `worse' than any other, merely performing comparably. Claiming improved performance would require significantly more work, and is largely irrelevant while the detrending methods are undergoing refinement, which is likely to happen for the duration of the K2 mission. The purpose of our method is to allow the rapid production of this catalogue so that it can be used reasonably soon after each data release.

\subsection{Lightcurve File Description}
The detrended light curve data provided with this catalogue is presented in FITS format \citep{Pence:2010jr}. We take the originally available target pixel files from the MAST archive, and add to these several additional data columns and headers. These are detailed in Tables \ref{tabnewheaders} and \ref{tabnewcolumns}. This allows all of the initial information provided by the K2 team to be preserved through the process. A description of the original files can be found at MAST.

\begin{table*}
\caption{Additional FITS Headers in primary extension}             
\label{tabnewheaders}      
\centering                          
\begin{tabular}{l r }        
\hline\hline                 
Label & Description \\    
\hline                        
    &  \\ 
 CPIX21& Coord of extraction, axis 2, segment 1\\
CPIX11 &Coord of extraction, axis 1, segment 1\\
AP\_RAD  & Squared radius of aperture. See Section \ref{sectextract} \\
SPLIT   & Time extraction split at (0 if no split used)\\
CPIX22  & Coord of extraction, axis 2, segment 2\\
CPIX12  &Coord of extraction, axis 1, segment 2\\
RA\_EXT  & RA of central extracted pixel, segment 1. Derived from supplied WCS data\\
DEC\_EXT & DEC of central extracted pixel, segment 1. Derived from supplied WCS data\\
DETNBINS  & Number of bins used in each axis for detrending (See Section \ref{sectdetrend})\\
\hline                                   
\end{tabular}
\end{table*}

\begin{table*}
\caption{Additional Data Columns (in file extension 1)}             
\label{tabnewcolumns}      
\centering                          
\begin{tabular}{l l r}        
\hline\hline                 
Label & Unit & Description \\    
\hline                        
APTFLUX  &  Counts & Extracted lightcurve\\ 
APTFLUX\_ERR  & Counts & Extracted lightcurve error \\ 
APTFLUX\_BKG  & Counts &  Calculated background\\ 
APTFLUX\_BKG\_ERR  & Counts & Calculated background error \\ 
DETFLUX  & Relative Flux &  Detrended lightcurve\\ 
DETFLUX\_ERR  & Relative Flux & Detrended lightcurve error \\ 
CENT\_ROW & Pixel coordinates & CCD Row Centroid \\
CENT\_COL & Pixel coordinates & CCD Column Centroid\\
\hline                                   
\end{tabular}
\end{table*}

\section{Catalogue}

\begin{table*}
\centering
\caption{Catalogue Table. An extract from the table is shown. The full table is available online.}
\label{tabcatalogue}

\begin{tabular}{@{}lllllr@{}}
\hline
EPIC ID & Type & Range & Period & Amplitude &Proposal Information\\
         &    & \%  &   d & \% & \\
\hline
\hline
:&:&:&:&: &: \\
201858862&AP&1.47&0.000000&0.00&\\
201859140&QP&69.11&0.483664&4.19&1018 (RR Lyrae)\\
201859398& P&4.08&2.656481&1.02&\\
201859496&AP&6.37&0.000000&0.00&1025 (AGN)\\
201859551&QP&3.67&9.374306&1.27&\\
:&:&:&:&: &: \\
\hline

\end{tabular}
\tablefoot{When online, clicking on an object ID will show detailed plots, as well as allow download of it's detrended light curve}

\end{table*}

The catalogue is presented in Table \ref{tabcatalogue}, and the full version can be found online\footnote{http://deneb.astro.warwick.ac.uk/phrlbj/k2varcat/}. The fields in the Table are described in Section \ref{sectfields}.

\subsection{Variable Detection and Classification}
\label{sectobjclass}
Once the detrended lightcurve for each object was available, we proceeded to search for variability. If the amplitude of the light curve (i.e. half the full range) was less than 3 times the median noise level the object was automatically discarded. For the remaining lightcurves a weighted, floating mean Lomb-Scargle (hereafter LS) periodogram \citep{Lomb:1976bo,Scargle:1982eu} with an oversampling factor of 4 was created, following the method of \citet{Press:1989hb}. Periods between 0.241 and 0.258 days were removed to avoid the 6-hour thruster firing timescale. These limits were determined through experimentation. Each periodogram, alongside the detrended and extracted lightcurves, was then classified by eye, and a period selected if appropriate. It is possible that in some cases classified variability in the catalogue is in fact due to systematic instrumental noise, although this was avoided when possible (for example, if many lightcurves shared the same variation). This problem is most apparent for longer period variation (greater than 20 days), and appears to be more common in Campaign 1 than Campaign 0.

After classification, period refinement was performed on objects marked as eclipsing binary (EB), periodic (P) or quasi periodic (QP). For P or QP lightcurves the LS periodogram was rerun with a higher oversampling factor of 20, over a range within $\pm 10$\% of the previously marked period. The most significant peak within this range is then given as the catalogue period. For EBs a Phase Dispersion Minimisation periodogram was created \citep{Stellingwerf:1978fe,SchwarzenbergCzerny:1997tq}, as these perform better on non-sinusoidal signals than do LS periodograms. This was run with a frequency resolution of $10^{-5}$($10^{-4}$ for objects with periods below 1d for efficiency reasons) over the same narrow range using 100 bins, and the most significant peak taken, in order to refine the only approximate LS period for the EB systems.

We imposed a limit of 15 days on the periods of objects within Campaign 0, due to the 32 day data baseline. For objects in Campaign 1, a limit of 20 days was imposed. The baseline for Campaign 1 (79 days) could allow longer periods, but due to the detrending method used, and specifically the splitting of the data into separate halves for detrending, signals on longer periods would not be robust. This does not apply to eclipses however, and as such no period limit was imposed on EB systems. Some EBs are classified without a period. In these cases either the period was too long to provide multiple eclipses, or for some other reason the period was uncertain. These generally represent the longest period objects in the catalogue, and so should be of special interest.

It is important to note that Campaign 0 data was classified before the WCS information (i.e. Data Release 2) was made available. At this time lightcurves were extracted using the brightest pixel in the central 9x9 box of each object's window. The brightest pixel was determined using the median average over time of each pixel in this box. Apertures were then placed centred on this brightest pixel, and given equal size for all targets. This size can be recreated using a value of AP\_RAD (see Section \ref{sectextract}) of 8. Therefore when a brighter object lies within \mytilde20" of the target object there is a significant chance that the classification was performed on the brighter object. Lightcurves are presented after extraction using Data Release 2 and the WCS information, with the implied reduced chance of blending. Classification was not repeated due to the significant time involved in performing it.

\subsection{Fields}
\label{sectfields}
\begin{enumerate}
\item \textbf{EPIC ID}\\
ID of target from the EPIC catalogue. Spans 201122454--210282474.
\item \textbf{Type}\\
 Lightcurves were classified by eye as Eclipsing Binary (EB), Periodic (P), Quasi-Periodic (QP), or Aperiodic (AP). Periodic classification implies a sinusoidal variation of constant period and amplitude. Quasi-periodic objects have amplitude or period variations, or a lightcurve non-sinusoidal in shape. Aperiodic objects showed no periodicity (though an object may also be classified as AP if it had no dominant periodicity). In many cases these objects may be periodic but with periods greater than 15 days for Campaign 0 or 20 days for Campaign 1, a limit imposed due to the data baseline (and the split used when detrending Campaign 1 data). Users should be aware that objects which should be classified as P can be misclassified as QP due to noise, and more rarely vice versa.
\item \textbf{Range}\\ 
The lightcurves were binned into 10 point wide bins and the median of each bin found. The range given is the maximum bin less the minimum, in flux units relative to the overall data median. In some cases outliers or remnant noise can affect this calculation, leading to ranges larger than are shown. Spans 0.03--429.85\%.
\item \textbf{Period}\\
The most significant peak from a Lomb-Scargle periodogram, for objects classified as P or QP. For eclipsing binaries the peak was found using Phase-Dispersion Minimisation (see Section \ref{sectobjclass}). Where possible the true period rather than an alias is given, even if the aliases were more significant. Zero for AP objects. No periods larger than 15 days for Campaign 0 and 20 days for Campaign 1 are shown to avoid spurious detections due to the data baseline. For the same reason, while we report periods up to these limits those within \mytilde 5 days of them should be treated with some caution. However, EB objects have no period limits imposed. Spans 0--59.889024 days.
\item \textbf{Amplitude}\\
The semi-amplitude of the lightcurve at the stated period, for objects classified P or QP. This was calculated through phase-folding the lightcurve, binning it into 40 evenly spaced bins, then taking the median of each bin. The semi-amplitude represents half of the maximum minus minimum bin value, in flux units relative to the overall data median. Short period objects will show reduced amplitude due to the cadence of the observations. For EB objects the number of bins was increased to 300, to improve detection of narrow eclipses. The resulting eclipse depth is then given directly (i.e. not halved as is the case for other objects). Zero for AP objects. Spans 0--96.78\%.
\item \textbf{Proposal Information}\\
Guest Observer proposals relating to the object. Only variable star related proposals are shown. If possible, the specific variable types which each proposal is related to are given in brackets.
\end{enumerate}

\section{Conclusion}
We have presented a catalogue of variable stars and eclipsing binaries from K2 Campaigns 1 and 0. This catalogue will be updated with each K2 data release, which we hope will provide a valuable resource for users interested in these objects. Detrended lightcurves for catalogue objects are also available, and compare favourably to already available detrending methods. We hope to make available detrended lightcurves for objects not found in the catalogue at a future date.

\begin{acknowledgements}
We thank the anonymous referee for a detailed review of the manuscript. The data presented in this paper were obtained from the Mikulski Archive for Space Telescopes (MAST). STScI is operated by the Association of Universities for Research in Astronomy, Inc., under NASA contract NAS5-26555. Support for MAST for non-HST data is provided by the NASA Office of Space Science via grant NNX13AC07G and by other grants and contracts. The authors would like to thank Thomas Marsh for use of his periodogram generating Python code.
\end{acknowledgements}


\bibliography{papers140415}
\bibliographystyle{aa}


\end{document}